\documentstyle[preprint,epsfig,aps]{revtex}
\begin{document}
\title{ELECTRON-ION STRUCTURE FACTORS AND THE GENERAL ACCURACY OF LINEAR 
RESPONSE}
\author{A.A. Louis} 
\address{ University Chemical Laboratories, 
Lensfield Rd, Cambridge CB2 1EW, UK}
\author{N.W. Ashcroft}
\address{ Cornell Center for Materials Research,
and Laboratory of Atomic and Solid State Physics,
Cornell University, Ithaca, NY 14853-2501}
\date{\today}
\maketitle
\begin{abstract}
 We show that electron-ion structure factors in 
fluid metallic systems can be well understood
from an application of {\em linear response} in the electron
system, combined with hard-sphere like correlation for the ionic component.
In particular, we predict that electron-ion structure factors fall 
into two general classes, one for high ($Z>3$) and one for low ($Z\leq2$)
 valence metals, and
make suggestions for experiments to test these ideas.  In addition, we
show how the general success of electronic 
linear response for most metallic systems 
stems in part from an intrinsic
interference between atomic and electronic length scales which weakens the
nonlinear response.  The main exception to this is metallic hydrogen. 
\end{abstract}
 \vspace{20pt}

\section{Introduction}

At a near fundamental level, liquid metals are complex binary fluids
consisting of ions in a sea of conduction electrons, their physical
properties linked to the three corresponding correlation functions:
$S_{II}(k), S_{eI}(k)$, and $S_{ee}(k)$\cite{Ashc78}.  We present a simple and
evidently  accurate analytic scheme to calculate electron-ion
structure factors ($S_{eI}(k)$) by combining a hard-sphere
approximation for the ionic structure with  a simple linear
response theory for the electrons.  These structure factors are now in
principle accessible experimentally through recent advances in both
neutron and x-ray scattering techniques.  Another route to effective
electron-ion interactions therefore opens, but now through the fluid
state.

 We also address the evident success of the linear approximation 
 by studying a related problem,
the density, $\rho^{ind}({\bf \vec k})$, of an initially uniform
electron gas induced by an embedded pseudo-potential, $v^{ps}({\bf \vec
k})$.  By comparing linear and second order response to full
(Kohn-Sham\cite{Kohn65}) non-linear response, we show that even though 
$v^{ps}({\bf \vec k})$ is not necessarily a small perturbation, 
the consequent response series converges term by term.
The non-linear terms are significantly reduced by an
interference between atomic and electronic length scales for most
metals, the main exception to this being hydrogen.

\section{Discussion}
\subsection{Electron-ion correlation functions}

The electron-ion-structure factor can be written as\cite{Ashc78}:
\begin{equation}\label{eq5}
S_{eI}(k) = \frac{1}{\sqrt{N_e N_I}}
 < \hat{\rho}_e^{(1)}({\bf \vec k}) \hat{\rho}_I^{(1)}({\bf \vec k}) > 
  = \frac{n(k)}{\sqrt{Z}} S_{II}(k),
\end{equation}
 where $\hat{\rho}_j^{(1)}(k)$ is the Fourier transform
of the one-particle density operator of component $j$,
$S_{II}(k)$ is the ion-ion structure factor and $n(k)$
  is identified as the
  pseudo-electron density, or pseudo-atom (of valence Z).  Thus  electron-ion
correlations can be described by convolving 
the pseudo-atom with the ionic correlations.  The ionic correlations
are themselves well described by a Percus-Yevick hard-sphere structure
factor\cite{Ashc66b}, while the pseudo-atom is described by a standard
linear response formulation:
\begin{equation}\label{eq5a}
n(k) = \chi_1(k) v^{ps}(k)
\end{equation} 
where $\chi_1(k)$ is the well known linear response function; to approximate
$\chi_1(k)$  we use a Local Density Approximation (LDA)
 local field factor\cite{Hafn87}.
 The electron-ion interaction is modeled by a simple local
one-parameter empty-core pseudopotential\cite{Ashc66} i.e.:  $v^{ps}(k) =
-(4 \pi e^2/k^2) cos(kR_c)$, where $R_c$ is the core radius
and the pseudo-potential goes through zero at $k_0 = \pi/2 R_c$. (We note that
at this linear level, the effects of ionic averaging on the pseudo-atom 
are ignored\cite{thesis}).
 Using the approximations above  in~(\ref{eq5}), we compare our 
approach in Fig. 1 to the {\em full ab-initio} Car-Parrinello\cite{Car85} calculations
of de Wijs {\em et al}\cite{deWi95}.  The correspondence is striking,
especially when we note that the parameters $\eta$ and $R_c$ are {\em
a priori} set by other physical properties (no fitting is necessary).

\begin{figure}
\begin{center}
\epsfig{figure=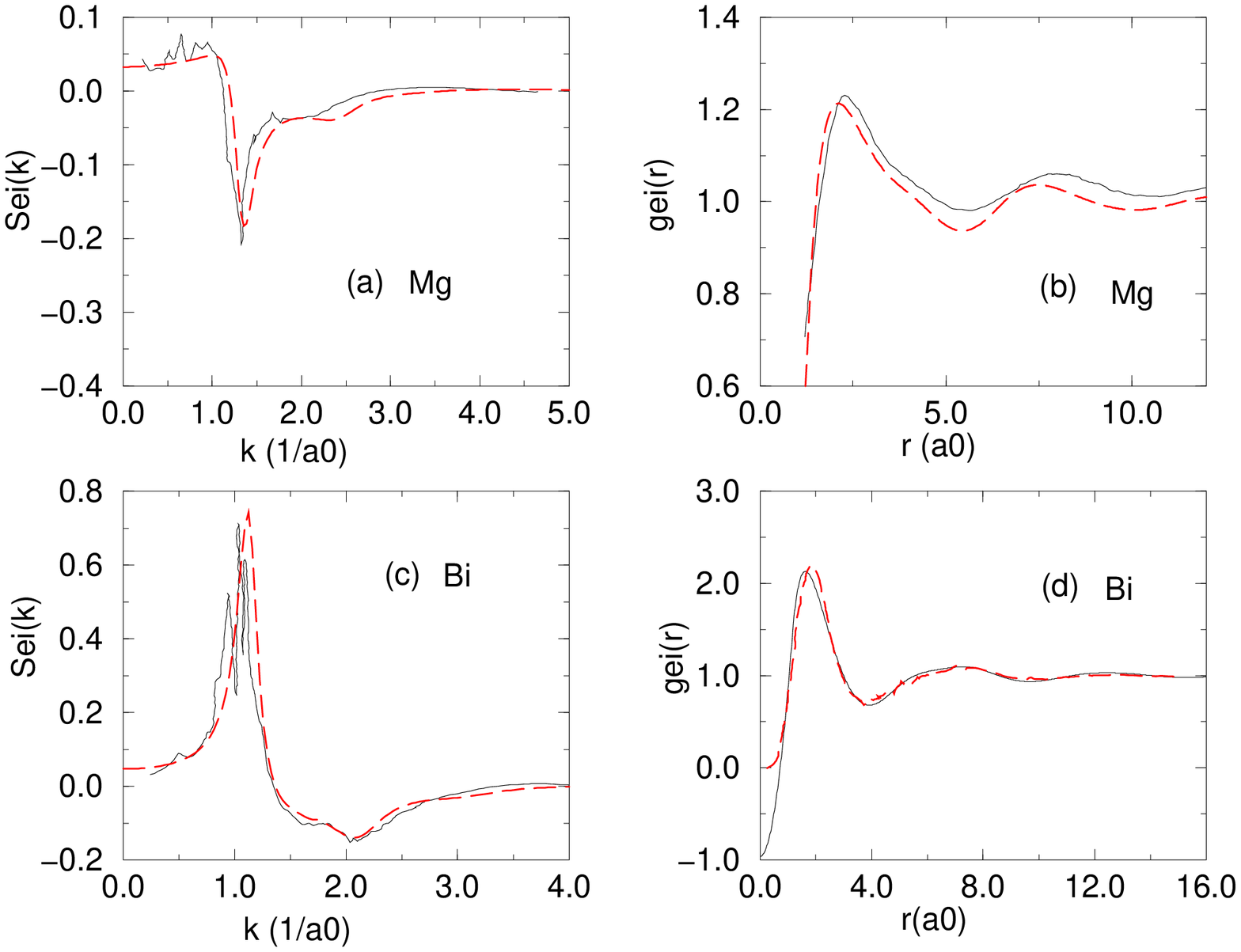,width=17cm}
\caption{}
\label{figfullBi_Mg}
\end{center}
\end{figure}

Besides a semi-quantitative description of electron-ion structure
factors, this linear response theory now provides an important
qualitative insight into the form of the electron-ion structure
factors\cite{Loui97b}.  The pseudo-atom density, $n(k)$, is typically
largest for smaller $k$ and thereafter 
rapidly declines for larger $k$, while the
near classical ion-ion structure factor, $S_{II}(k)$, follows an inverse
behavior; it is small for small $k$.  Together with the product
form~(\ref{eq5}) this implies that the shape of the electron-ion
structure factor, $S_{eI}(k)$, is determined primarily by the the
position of the {\em zero-crossing}, $\bar{k}_0$, of $n(k)$ with respect
to the {\em first maximum}, $k_p$, of $S_{II}(k)$.  If $\bar{k}_0 <
k_p$, then $S_{II}(k)$ selects (or filters) the negative part of
$n(k)$ and $S_{eI}(k)$ takes a form similar to that of Mg (Fig.  1
(a)).  Conversely, if $\bar{k}_0 > k_p$, then the ion-ion structure
factor selects (or filters) the positive part of $n(k)$, and again,
$S_{eI}(k)$ takes a form similar to that of Bi (Fig.  1
~(\ref{figfullBi_Mg} (c)).  Since $\chi_1(k)$ is positive definite,
the zero-crossing in linear response occurs at $k_0$.  The large slope
of $n(k)$ near the zero-crossing then implies that non-linear
corrections {\em must} have a small effect on the location of the
zero-crossing, and this,  together with the expected accuracy of linear
response,  implies that $\bar{k}_0 \sim k_0$.  For most metals,
$k_0$ is just a little less than $2k_F$, and the latter's ratio to
$k_p$ is well known:  for low valence $(Z \leq 2)$, $2k_F < k_p$;
for high valence $(Z\geq 3)$:  $2k_F > k_p$\cite{Zima72}.  This
accounts in a straightforward way for the two separate forms found by
deWijs {\em et al}\cite{deWi95}:  For Mg, $\bar{k}_0 < k_p$ $(Z=2)$,
which belongs to the {\bf low valence class} of electron-ion structure
factors.  For Bi, $\bar{k}_0 > k_p$ $(Z=5)$ and we may refer to this
as the {\bf high valence class} of electron-ion structure
factors\cite{deWijssei}.  Generally ions of valence $Z \leq 2$ belong
to the low valence class while ions with valence $Z > 3 $ belong to
the high valence class.  Ions with valence $Z=3$ typically belong to
the high valence class also, although they may be characterized by a
crossover form\cite{thesis}.  The analytical approach above can easily
be extended by using the modern theory of classical liquids 
  to obtain
improved ion-ion structure factors\cite{Cusa76}, but to include second
order contributions to the pseudo-atom $n(k)$ necessitates not only
second order electron response, but also contributions from ion-ion
triplet structure.  The latter can  be carried out with concepts from
the theory of classical liquids\cite{thesis}.

\subsection{Proposed Experiments}

The principal features of electron-ion structure factors can be
measured by exploiting the differences between   the 
x-ray scattering structure factor, $S_{II}^X(k)$, determined with 
a free-atom form factor, $f_A(k)$, and the structure factor, $S_{II}^N(k)$,
determined by neutron scattering\cite{Egel74,Chih87}.  As emphasized
by Chihara\cite{Chih87}  the x-ray
structure factor for liquid metals will equal $S_{II}^N(k)$ only when
  determined with an 
ionic form factor augmented by a  pseudo-atom form factor; i.e.
$f_I(k) + n(k)$, so that:
\begin{equation}\label{eq10}
\frac{S_{II}^X(k)}{S_{II}^N(k)} = \frac{|f_I(k) + n(k)|^2}{|f_A(k)|^2}.
\end{equation}
 This effect is 
 clearly expected to be largest for metals with larger ratios of
valence to core electrons. Thus we predict a small effect for metals
with smaller valence to core ratio such as Na or K, 
 a 2\% difference at the 1st peak of the structure for Li (ratio$=1:2$)
 or Al (ratio$=3:10$), but by far the largest effect  for Be (ratio $=1:1$)
 where the difference  at the
principal peak of the structure factor could be as high as 7\%, well
within experimental range.  Another interesting candidate would be
metallic Si (ratio $=4:10$) since covalent effects still make
themselves felt in the liquid state suggesting that experiments 
could reveal effects beyond linear response.

To date the experimental electron-ion structure factors and related 
pseudo-atoms show
considerably more  structure than indicated by theoretical
predictions\cite{Take85}. 
Significant experimental challenges are faced 
in the
accuracy resulting from subtraction of two sets of data obtained by quite
different means, each with important (but different) systematic corrections; 
however the present approach suggests that the current differences between
x-ray and neutron scattering should be reexamined (see also\cite{Chih87}). 
Using a pseudo-atom instead of the full free atom as
a form factor can assist in comparing neutron and x-ray measurements
and help unravel various systematic corrections  applied.
The advent of high precision x-ray and neutron sources currently
coming on-line suggests that these proposed effects can be systematically
explored.

\subsection{Non-linear response of an atom in an electron gas}

The evident  (and long-standing) success of the linear response approximation 
for electron response\cite{Hafn87}, here demonstrated for
electron-ion structure factors, calls for further investigation.  The
accuracy of linear response in a crystalline solid is commonly
attributed to the fact that the structure dependent reciprocal lattice
vectors are typically near the pseudo-potential zero-crossing, $k_0$,
with the associated inference that the net scattering is smaller than one would
naively expect\cite{Hafn87}.  For
liquids or other disordered systems such arguments are less appropriate.  To
examine the strength of linear response in the absence of  ionic structure,
 we consider a simpler problem, namely the
response of the interacting electron gas to a single ion, where the
electron-ion interaction is modeled by a simple local one-parameter
empty-core pseudopotential\cite{Ashc66}.  We will compare two routes
to the induced density, $\rho^{ind}(k)$.  The first follows from
solving the Kohn-Sham equations\cite{Kohn65} exactly (for the given pseudopotential)
 within the local density
approximation (LDA), the second from the standard expansion
of the 
response in powers of the perturbing (pseudo)potential, i.e.;
\begin{equation}\label{eq1}
 \rho^{ind}(k) = \chi_1(k) v^{ps}(k) +
\sum_{\vec{k}_1} \sum_{\vec{k}_2}
\chi_2(k,k_1,k_2)v^{ps}(k_1)v^{ps}(k_2) + \cdots \, .  
\end{equation}
Here the response functions, $\chi_n(k_1...)$, are properties of the
{\em homogeneous} interacting electron gas. In Fig. 2 we compare
the explicit second order response with an LDA local field 
factor\cite{thesis,Lloy68}  to the full non-linear LDA response.
 Clearly the 
non-linear response is well characterized by the 2nd order term,
implying that the success of linear response is not due to
cancellation between higher order terms of opposite sign but instead
that each successive term is individually small compared to the previous
term in the expansion; {\em the response series
converges very rapidly, term by term}.

The non-linear response is largest for atomic parameter $R_c =0$ (hydrogen), and
decreases with a larger atomic-parameter, $R_c$, as might be physically
anticipated.  However as $R_c$ increases from zero, a noticeable secondary
minimum occurs when the inverse atomic length, $k_0$, is equal to $2k_F$.  For
the cases plotted in Fig.  2, the maximum in second order response at $k_0/2k_F
= 1$ (or $R_c/r_s = 0.41$) is reduced by an entire {\em order of magnitude} when
compared with the maximum in second order response calculated for hydrogen
($R_c/r_s = 0$), and is typically equal to the value at $3 R_c/r_s$.  The
physics behind this minimum is attributed to the following; the second order
response function, $\chi_2(k,k_1,k_2)$, peaks when the summed arguments
in~(\ref{eq1}) are close to $2k_F$\cite{thesis}.  If the pseudo-potential
zero-crossing, $k_0$, is near the response peaks at $2k_F$, a maximal
cancellation or {\em maximal destructive interference of the atomic and
electronic length scales} occurs, leading to a minimum in second order response.
The ratio of the atomic and electronic length scales is set primarily by the
volume energy terms in the total ground state energy, and is almost independent
of structure\cite{Hafn87}; $k_0/2k_F$ lies between $0.75$ and $1$ for most
metals, and is therefore very close to the secondary minimum in the non-linear
response.

As noted, the effect we discuss originates from an {\em interference} between
intrinsic {\em atomic} and {\em electronic} length scales, but it also
complements the argument given for crystalline solids alone, which stems from
the {\em confluence } of an {\em atomic} and a {\em structural} length scale.
The clear exception to these interference effects is again the singular case of
a point-charge $(v^{ps}(k) \sim 4 \pi e^2/k^2)$, i.e.  the case of hydrogen,
which has no well-defined core-length scale, $k_0$, no oscillations in the
potential and thus no interference effect in the higher order terms.  In
contrast to other systems, non-linear response terms are large {term by term}.
In fact, the response series may not even formally converge and care must be
taken when applying concepts derived from linear-response theory to hydrogen (it
is not a simple material).

Finally we note that the second
 order response contribution is of the same order as
the difference between first   order response with or without local field
corrections.  In addition, the combined effects of exchange and correlation
partially cancel between first and second order, implying that {\em neglect of
higher order response results in an over-estimation of the role of exchange and
correlation}, which, in turn, has important implications for the widespread
application of linear response theory in the derivation of effective ion-ion
potentials in (simple) metals.

\section{Conclusions}

The electron-ion structure factors of liquid simple-metals 
are well described by a
 simple linear response theory augmented by linear response for the electrons.
 This approach suggests two main classes of electron-ion correlation functions,
 one for high and one for low valence metals.  Experimental advances in x-ray
 and neutron-scattering may be able to provide measurements of these
 electron-ion correlation functions, with liquid Be being the most promising
 candidate.  A route to information on fundamental electron-ion interactions
 therefore becomes available through the fluid state.  Finally, the well
 documented success of the linear response approximation for electrons stems in
 part from an interference effect between atomic and electronic
 length scales.

This work was supported by the NSF through the Cornell Center for 
Materials Research under Grant No.  DMR96-32275.  We
especially thank Professor Karsten  Jacobsen for making an 
LDA Kohn-Sham program
available to us, and Dr David Muller for helpful suggestions.

\newpage
\begin{center}
{\bf \underline{FIGURE CAPTIONS} }
\end{center}

{\bf FIGURE 1}

The electron-ion structure factors $S_{eI} (k)$ and related electron-ion
correlation functions $g_{ei}(r)$ for Mg and Bi:  Car-Parrinello results of de
Wijs {\em et al}{\protect\cite{deWi95}} (solid line) vs.  the simple
linear-response approach augmented by a hard-sphere approximation (dashed line).
Panel (a) shows $S_{eI}(k)$ and panel (b) shows $g_{eI}(r)$ for liquid Mg.
Panel (c) shows $S_{eI}(k)$ and panel (d) shows $g_{eI}(r)$ for liquid Bi.  For
Mg the parameters (taken from the literature) are:  $r_s$$=$$2.66a_0$,
$R_c$$=$$1.31a_0$ and for Bi the parameters (taken from the literature) are:
$r_s$$=$$2.25a_0$ and $R_c$$=$$1.15a_0$.  Both have a hard-sphere parameter,
$\eta$$=$$0.46$.  (note that for the $g_{eI}(r)$ the region inside the core
radius is not physically significant.)\vspace*{2cm}

{\bf FIGURE 2}

A comparison of full non-linear LDA response $[\rho(k)$$-$$\rho^{(1)}(k)]$
(solid line) to second order LDA response (dashed line) for an empty core
pseudo-potential with $R_c$$=$$1.5a_0$ embedded in an electron gas with density
parameter, $r_s$$=$$3a_0$.  For the scale, compare this to the full response with
the limit $\rho(k \rightarrow 0)$$=$$1$.  The higher order response is of the order of a
few $\%$ of the full response and in turn, the second order response captures
almost all the non-linear response.  (The small difference at $k\rightarrow 0$
is a numerical artifact stemming from the use of a large but finite real-space
cut-off radius in the Kohn-Sham procedure.)  In the insert is plotted the
maximum of the 2nd order response vs.  $R_c/r_s$ for $r_s$$=$$2a_0$(dotted),
$r_s$$=$$3a_0$(solid) and $r_s$$=$$5a_0$(dashed).  Note especially the minimum
at $R_c/r_s$$=$$0.41$ which corresponds to $k_0$$=$$2k_F$.  It is reduced by an
order of magnitude from the value at $R_c$$=$$0$ (hydrogen) and is traced to an
interference between atomic and electronic length-scales.

\newpage

\vglue-.5cm

\begin{figure}
\begin{center}
\epsfig{figure=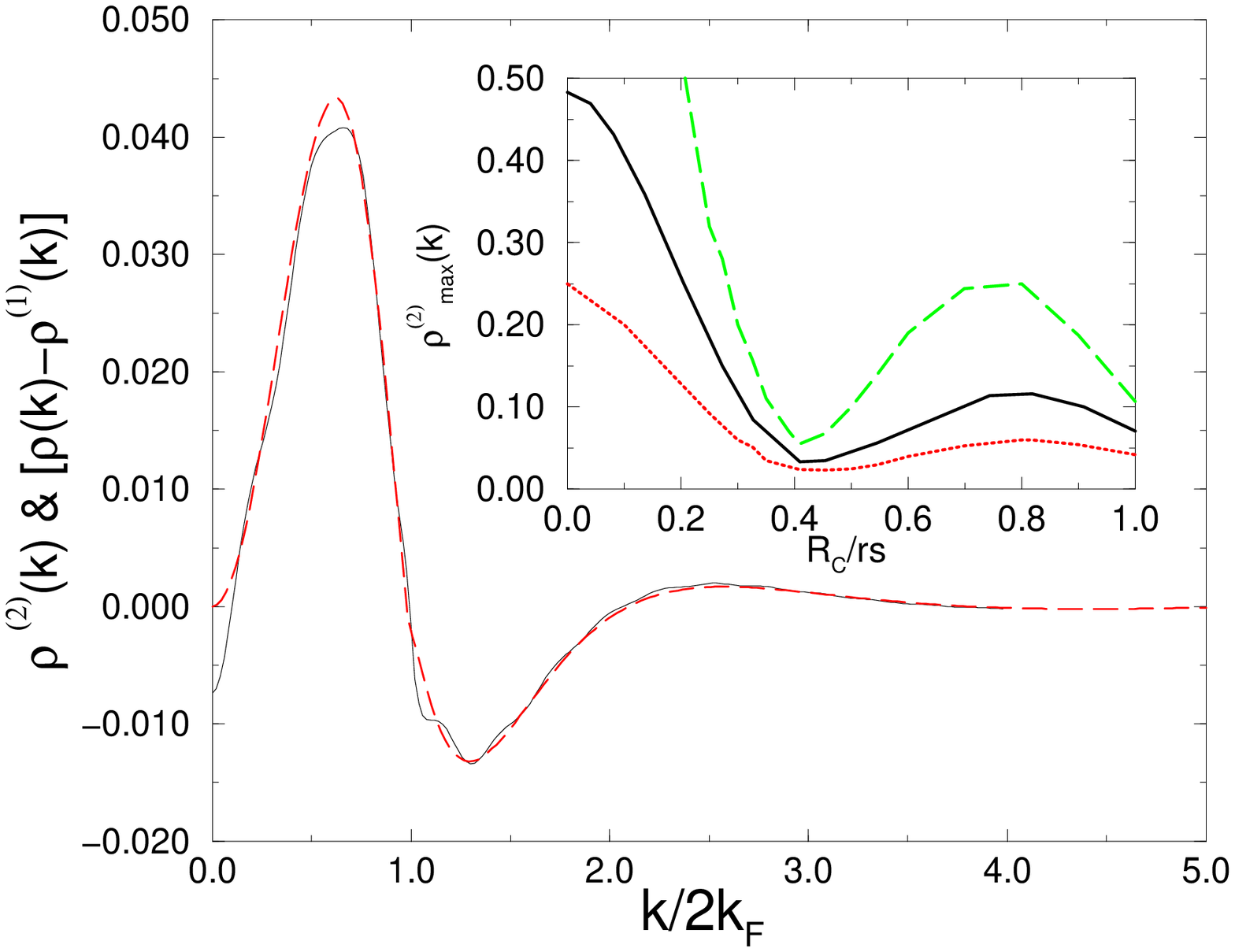,width=17cm}
\caption{}
\label{fig4.1.5}
\end{center}
\end{figure}
\vglue-.5cm


\begin{thebibliography}{99}
\bibitem{Ashc78} N.W. Ashcroft and D. Stroud, Solid State Physics {\bf
33}, 1 (1978).
\bibitem{Kohn65}W. Kohn and L.J.  Sham, Phys. Rev. {\bf 140}, A1133
(1965).
\bibitem{Ashc66b} N. W. Ashcroft and J. Lekner, Phys. Rev. {\bf 165},
83 (1966).
\bibitem{Hafn87} J. Hafner, {\em From Hamiltonians to Phase Diagrams},
(Springer Verlag, Berlin, (1987)).
\bibitem{Ashc66} N.W. Ashcroft, Phys. Lett. {\bf 23}, 48 (1966).
\bibitem{thesis} A.A. Louis,  {\em
  Quantum Dissipation from Phonons; Metallic Hydrogen; Electron-Ion Mixtures},
  PhD Thesis,     Cornell University (1997) 
\bibitem{Car85} R. Car and M. Parrinello, Phys. Rev. Lett. {\bf 55 }, 2471
   (1985).
\bibitem{deWi95} G. A. de Wijs,  B. Pastore, A. Selloni, and W. van
der Lugt, Phys. Rev.  Lett. {\bf
75}, 4480 (1995).
\bibitem{Loui97b} A.A. Louis and N.W. Ashcroft,   Phys. Rev. Lett.
{\bf 81}, 4456 (1998).
\bibitem{Zima72} J. M. Ziman, {\em Principles of the Theory of
Solids, 2nd Ed}, (Cambridge University Press, Cambridge (1972)), p 227.
\bibitem{deWijssei} 
In their interesting paper de Wijs {\em et.  al.}\cite{deWi95} suggest that the
difference between Mg and Bi electron-ion structure factors is partially due to
the nearly free electron (NFE) bonding of $l$-Mg as against the remnant
covalency of $l$-Bi.  However a high valence NFE metal such as $l$-Pb should
have the same form of $S_{eI}(k)$as $l$-Bi.  
\bibitem{Cusa76} S. Cusack, N.H. March, M. Parrinello and M. P. Tosi,
J. Phys F: Met. Phys. {\bf 6}, 749 (1976).
K. Hoshino and M. Watabe, J. Phys. Soc. Japan {\bf
61}, 1663 (1992).
\bibitem{Egel74} P.A. Egelstaff, N.H. March and N.C. McGill, Can. J.
Phys. {\bf 52}, 1651 (1974).
\bibitem{Chih87} J. Chihara, J. Phys. F: Met. Phys. {\bf 17}, 295 (1987).  See also 
 J.A. Anta, B.J. Besson, and P.A. Madden Phys. Rev. B. {\bf 58}, 6124 (1998). 
\bibitem{Take85} S. Takeda, S. Tamaki and Y. Waseda, J. Phys. Soc.
Japan, {\bf 54} 2552 (1985); S. Takeda, S. Harada, S. Tamaki and Y.
Waseda, J. Phys. Soc.
Japan, {\bf 55} 184 (1986); {\em ibid} {\bf 55}  3437 (1986),
 {\em ibid}, {\bf 58}, 3999 (1989), {\em ibid} {\bf 60}, 2241 (1991),
{\em ibid} {\bf 63} 1794, (1994).
\bibitem{Lloy68} P. Lloyd and C. Sholl J. Phys. C {\bf 1}, 1620
(1968).

\end{thebibliography}
\end{document}